\documentclass[12pt]{article}
\usepackage{amsmath,bbm,amssymb,amsfonts,epsfig}
\usepackage{longtable}
\usepackage{multirow}
\usepackage{amsfonts}
\usepackage[english]{babel}

\newcommand {\slsh} [1] {\not{\hbox{\kern-2pt${#1}$}}}

\newcommand {\beq} {\begin{equation}}
\newcommand {\eeq} {\end{equation}}
  \newcommand {\ber}{\begin{eqnarray*}}
  \newcommand {\eer} {\end{eqnarray*}}
\newcommand {\bea}{\begin{eqnarray}}
\newcommand {\eea} {\end{eqnarray}}

\newcommand{\Dslash}{\,{\raise.15ex\hbox{/}\mkern-12mu D}}

\newcommand{\point}{\;.}
\newcommand{\comma}{\;,}

\newcommand{\tr}{\mathrm{tr}}
\newcommand{\Tr}{\mathrm{Tr}}
\newcommand{\paths}{\mathcal{C}}

\newcommand{\Adj}{\textrm{Adj}}
\newcommand{\Asym}{\textrm{As}}
\newcommand{\Sym}{\textrm{S}}
\newcommand{\Repr}{\textrm{R}}

\begin{document}

\title{\bf Lattice Study of Planar Equivalence:\\ The Quark Condensate}
\author{\large Adi Armoni, Biagio Lucini, Agostino Patella\\
~\\
Department of Physics, Swansea University\\
Singleton Park, Swansea, SA2 8PP, UK\\
~\\
\and Claudio Pica \\
~\\
Physics Department, Brookhaven National Lab\\
Upton, NY 11973-5000, USA
}

\date{}

\maketitle
~\\
\vspace{-12cm}
~\\
\begin{flushright}
BNL-NT-08/12\\
\end{flushright}
~\\
\vspace{8cm}
~\\

\abstract{
We study quenched $SU(N)$ gauge theories with fermions in the two-index symmetric, antisymmetric and the adjoint representations. Our main motivation is to check whether at large number of colours those theories become non-perturbatively equivalent. We prove the equivalence assuming that the charge-conjugation symmetry is not broken in pure Yang-Mills theory. We then carry out a quenched lattice simulation of the quark condensate in the symmetric, antisymmetric and the adjoint representations for $SU(2)$, $SU(3)$, $SU(4)$, $SU(6)$ and $SU(8)$. We show that the data support the equivalence and discuss the size of subleading corrections.
}


\section{Introduction}
\label{sec:introduction}

\noindent

It was proposed recently that an $SU(N)$ gauge theory with a Dirac fermion in the symmetric/antisymmetric representation becomes equivalent to a theory with a Majorana fermion in the adjoint representation, in the large-$N$ limit \cite{Armoni:2003gp,Armoni:2004ub}. A necessary and sufficient condition for the equivalence to hold is the non-breaking of charge conjugation symmetry \cite{Kovtun:2004bz,Unsal:2006pj} (this condition is equivalent to the absence of closed string tachyons in the string realisation of the theory \cite{Armoni:2007jt}). Whereas charge conjugation is spontaneously broken when the theory is compactified on a small circle \cite{Unsal:2006pj,Barbon:2005zj}, it is expected to be restored as the size of the circle increases above a critical value \cite{Hollowood:2006cq,DeGrand:2006qb,Lucini:2007as}, in a mechanism similar to the deconfinement/confinement transition. Thus 'planar equivalence' is expected to hold on $R^4$.

The equivalence between a theory with a massless fermion in the adjoint representation (a supersymmetric theory) and a non-supersymmetric theory enables to copy analytic non-perturbative results from the former to the latter. In particular, the equivalence predicts the value of the quark condensate in the non-supersymmetric theory. Of prime interest is the theory with a fermion in the antisymmetric representation, since it becomes one-flavor QCD for $SU(3)$. Therefore, it is possible to estimate the quark condensate in one-flavor QCD, by copying its value from the SUSY theory \cite{Armoni:2003fb,Armoni:2003yv}. Planar equivalence is reviewed in Refs. \cite{Armoni:2004uu,Armoni:2007vb}. Refs. \cite{Patella:2005vx,DeGrand:2006uy,KeithHynes:2007vs,Farchioni:2007dw,DelDebbio:2008wb} contain related lattice papers.

In this paper, we would like to make a first attempt at checking the proposed planar equivalence by lattice calculations. In particular we are interested here in the quenched theories. Since the vacuum of the pure Yang-Mills preserves charge-conjugation symmetry, the orientifold planar equivalence is expected to be valid in this case. Moreover numerical simulations are quite cheap for the quenched theories. Nevertheless, the quenched theories offer an interesting benchmark for studying the subleading corrections. Therefore, we carry out a quenched simulation of the quark condensate in the symmetric, the antisymmetric and the adjoint representations, for $SU(N)$ with various values of $N$:
\begin{subequations}
\begin{flalign}
& \langle \bar \psi \psi \rangle _{\Sym / \Asym} = \int DA _\mu (\exp i S_{\rm YM} ) \, \Tr \frac{1}{i \gamma_\mu D_{\Sym / \Asym \, \mu} - m}\, ,\\
& \langle \lambda \lambda \rangle _{\Adj} = \int DA _\mu (\exp i S_{\rm YM} ) \, \Tr \frac{1}{i \bar{\sigma}_\mu D_{\Adj \, \mu} - m}\, ,
\end{flalign}
\end{subequations}
with $D_{\Repr \mu} = \partial _\mu -i A_\mu ^a T^a _\Repr$ and $m$ is the quark mass (henceforth $\psi$ and $\lambda$ represent respectively a Dirac and a Majorana fermion). Since the trace of the quark propagator can be expanded in loops, the condensate in the various representations arises from the planar and non-planar quark loops depicted in Figs.~\ref{adjointfig} and~\ref{symmfig} below.

\begin{figure}[ht]
\centerline{\includegraphics[width=2cm]{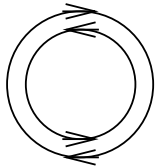}}
\caption{\footnotesize The 't Hooft notation for a quark loop in the adjoint representation. } \label{adjointfig}
\end{figure}

\begin{figure}[ht]
\centerline{\includegraphics[width=5cm]{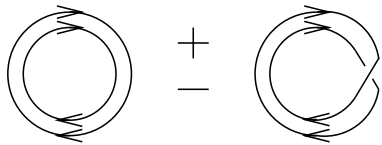}}
\caption{\footnotesize The 't Hooft notation for a quark loop in the symmetric/antisymmetric representation.} \label{symmfig}
\end{figure}

In particular, the graph~\ref{symmfig} suggests the following $N$ dependence for the quenched quark condensate in the symmetric/antisymmetric representations:
\begin{subequations} \label{eq:intro:s/as}
\begin{flalign}
& \frac{ \langle \bar{\psi} \psi \rangle_{\Sym} }{N^2} = f\left( \frac{1}{N^2},m \right) + \frac{1}{N} g\left( \frac{1}{N^2},m \right) \comma \\
& \frac{ \langle \bar{\psi} \psi \rangle_{\Asym} }{N^2} = f\left( \frac{1}{N^2},m \right) - \frac{1}{N} g\left( \frac{1}{N^2},m \right) \point
\end{flalign}
\end{subequations}
This decomposition will be proved in Sect.~\ref{sec:quenched:condensate:corrections} and is very useful for disentangling the even and odd corrections from the lattice data.

Planar equivalence implies
\beq 
\lim _{N \rightarrow \infty} \left[ \frac{\langle \lambda \lambda \rangle_\Adj}{N^2}  -  \frac{\langle \bar \psi \psi \rangle _\Sym +  \langle \bar \psi \psi \rangle _\Asym }{2 N^2} \right] = 0 \, ,
\eeq
at any finite $m$. We shall show that lattice measurements of the fermionic condensates for $N = 2, 3, 4, 6, 8$ support planar equivalence, and that the first subleading correction describes quite well the results for the adjoint representation, while this is not true for the symmetric/antisymmetric representations.

The organisation of the paper is as follows. We begin by describing the lattice setup for the analytical proof and the numerical calculations~(Sect.~\ref{sec:lattice}). In Sect.~\ref{sec:quenched:condensate:equivalence} we provide an analytic proof of planar equivalence in the quenched approximation. In Sect.~\ref{sec:quenched:condensate:corrections} we prove Eqs.~\eqref{eq:intro:s/as}. Then, in Sect.~\ref{sec:calculation} we will present our numerical results (which are reported in the Appendix). A discussion of our results and an outlook on future developments will be presented in Sect.~\ref{sec:conclusions}.

\section{The lattice setup}
\label{sec:lattice}
In order to prove the equivalence between the condensates in the two-index representations, a regularisation scheme needs to be used. To make contact with the lattice calculations, we will provide a proof that uses the lattice as a regulator. However, we stress that although the details of the proof depend on the regularisation scheme, the end result is general and can be easily proved also in different schemes.

We consider a four-dimensional spacetime lattice with dimensions $L_i = N_i a$, where $a$ is the lattice spacing. Lattice points are labelled by $x \equiv (n_0,n_1,n_2,n_3)$, with  $n_i = 0, \ \dots, N_i -1$. Fermion fields live on the points of this lattice, while gauge fields live on the links. Boundary conditions are periodic for all fields in the spatial directions and for gauge fields in the temporal direction, antiperiodic for fermions in the temporal direction.

The quenched theory is described by the partition function
\begin{eqnarray}
\label{zeta}
Z = \int \left( {\cal D} U \right) e ^{ - S} \ ,
\end{eqnarray}
where $S$ is the lattice action. For $S$, we chose the Wilson action:
\begin{eqnarray}
S = \beta \sum_{x,\mu > \nu} \left( 1 - \frac{1}{N}{\cal R}\mathrm{e}\,\tr{U_{\mu \nu}(x)}\right) \ , \qquad \beta = 2N/g^2 \ ,
\end{eqnarray}
where $U_{\mu \nu}$ is the parallel transport of the link variables $U_{\rho}(z) \in SU(N)$ around the elementary square of the lattice identified by the point $x$ and the positive directions $\hat{\mu}$ and $\hat{\nu}$ ({\em plaquette}). The
integration measure ${\cal D} U$ in Eq.~\eqref{zeta} is the product of the $SU(N)$ Haar measures for each link variable.

For a theory with dynamical fermions, the fermionic part is described by the action
\begin{eqnarray}
S_F = \sum_{i,x,y} \bar{\psi}_i(x) D_{xy} \psi_i(y) \ ,
\end{eqnarray}
with $\psi_i$ the fermion field (a Grassmann variable whose dimension depends on the fermion discretisation and the colour group) for the $i$-th flavor and $D$ the Dirac operator (which also depends on the fermion discretisation chosen). In this work, fermionic observables are derived from the staggered Dirac operator in the generic representation $R$
\begin{eqnarray}
D_{xy} &=& m \delta_{xy} + K_{xy} = \nonumber \\
&=& m \delta_{xy} + \frac12 \sum_{\mu} \eta_\mu(x) \left\{ R[U_\mu(x)] \delta_{x+\hat{\mu},y} - R[U_\mu(x-\hat{\mu})]^\dagger \delta_{x-\hat{\mu},y} \right\} \comma
\end{eqnarray}
where $x$ and $y$ are two lattice points and $m$ is the bare quark mass in units of the lattice spacing. The $\eta_{\mu}(x) = (-1)^{\sum_{i=0}^{\mu - 1} n_i}$ are the Kawamoto-Smit phases, the staggered equivalent of the $\gamma$ matrices.

Staggered fermions have been chosen over alternative discretised fermions because in this formulation the Dirac operator is ultra-local and retains a remnant of chiral symmetry that protects the mass from additive renormalisation. Potential drawbacks of the staggered approach to lattice fermions (namely flavor doubling and mixing) are harmless for the problem we are investigating. In particular, flavour doubling (i.e. the fact that one lattice flavour corresponds to four continuous flavours) translates into a simple multiplication by four of the chiral condensate, since the theory is quenched. Although peculiarities deriving from the lattice formulation are not difficult to account for, one should keep in mind that at finite lattice spacing some properties of the lattice theory (like e.g. the degeneracy of the eigenvalues) are determined by the specific form of the Dirac operator. 

In this work, we will use one staggered flavour, described by the scalar Grassmann variable $\psi(x)$. The chiral condensate is given by
\begin{eqnarray}
\langle \bar{\psi} \psi \rangle_{\mathrm{YM}} = \frac{1}{V} \langle \Tr D^{-1} \rangle_{\mathrm{YM}} = \frac{1}{Z V} \int \left( {\cal D} U \right) \left( \sum_{x} \tr D^{-1}_{x,x} \right) e^{- S} \ ,
\end{eqnarray}
where $V = \prod_i L_i$ is the lattice volume, $D^{-1}$ is the inverse staggered Dirac operator, the subscript $\mathrm{YM}$ stresses that the theory is quenched (henceforth we will omit it), $\Tr$ is the trace over both the space and colour indices, and $\tr$ is the trace over the colour index.

Having defined the notations, we move to the proof of the equivalence between the condensate in the (anti)symmetric and in the adjoint representations in the large-$N$ limit.

\section{Analytical equivalence}
\label{sec:quenched:condensate:equivalence}
In the quenched theories the orientifold planar equivalence of purely gluonic observables is trivial. We want to investigate analytically the equivalence between fermionic observables, in particular the eigenvalue probability distribution (EPD) of the Dirac operator, and the fermionic condensate. The strategy is to expand the fermionic observables in Wilson loops, and to use the following relationship\footnote{From algebraic relationships:
\begin{subequations}
\begin{flalign}
& \tr \Sym[U] = \frac{ ( \tr U )^2 + \tr (U^2) }{2} \comma \\
& \tr \Asym[U] = \frac{ ( \tr U )^2 - \tr (U^2) }{2} \comma \\
& \tr \Adj[U] = | \tr U |^2 -1 \point
\end{flalign}
\end{subequations}
In the planar limit only the first term of each r.h.s. survives. Using the factorisation of the expectation values in the planar limit and the charge-conjugation invariance of the Yang-Mills vacuum, we get:
\begin{equation}
\frac{1}{N^2} \langle \tr \Sym[U] \rangle = \frac{1}{N^2} \langle \tr \Asym[U] \rangle = 
\frac{\langle \tr U \rangle^2}{2N^2} = \frac{|\langle \tr U \rangle|^2}{2N^2} =
\frac{1}{2N^2} \langle \tr \Adj[U] \rangle \qquad \textrm{as } N \rightarrow \infty \ .
\end{equation}
}:
\begin{equation}
\lim_{N \rightarrow \infty} \frac{1}{N^2} \langle \tr \Sym[U] \rangle = \lim_{N \rightarrow \infty} \frac{1}{N^2} \langle \tr \Asym[U] \rangle = \lim_{N \rightarrow \infty} \frac{1}{2N^2} \langle \tr \Adj[U] \rangle \comma
\label{eq:groupequivalence}
\end{equation}
in order to prove the equivalence. Although any regularisation of the theories could be used, the lattice is very convenient since the set of the closed paths is discrete. The expansion of the fermionic observable in Wilson loops should be written in general as a worldline path integral~\cite{Strassler:1992zr,D'Hoker:1995ax,D'Hoker:1995bj}. On the lattice these path integrals are replaced by discrete infinite sums and the issue of the convergence is better defined.

Let us begin with the analysis of the EPD of the Dirac operator. The eigenvalues of the staggered Dirac operator are of the form $m+i\lambda_\alpha$ (where ${\lambda_\alpha}$ are real numbers) and appear as pairs of complex conjugate values. The operator $H=D^\dagger D=m^2-K^2$ has doubly-degenerate eigenvalues of the form $m^2+\lambda_\alpha^2$; therefore, it is positive definite. Clearly, assigning the spectrum of $D$ is completely equivalent to assign the spectrum of $H$, once the mass has been fixed. Since it is more convenient to deal with real positive eigenvalues, in what follows we will refer to the eigenvalues of $H$ rather than $D$.

The EPD of $H$ is defined as:
\begin{equation}
\rho_\Repr(\tau) = \frac{1}{d_\Repr V} \sum_\alpha \langle \delta(\tau - m^2 - \lambda_\alpha^2[U]) \rangle \comma
\end{equation}
where $d_\Repr$ is the dimension of the representation $\Repr$, and $d_\Repr V$ is the total number of eigenvalues of the staggered Dirac operator. It is also useful to consider the Laplace transform of the EPD:
\begin{equation}
\hat{\rho}_\Repr(z) = \int \rho_\Repr(\lambda) e^{-z\lambda} d\lambda = \frac{1}{d_\Repr V} \sum_\alpha \langle e^{-z \left( m^2 + \lambda_\alpha^2[U] \right) } \rangle = 
\frac{e^{-z m^2}}{d_\Repr V} \langle \Tr e^{z K^2} \rangle \comma
\end{equation}
where $\Tr$ is the trace over both the site and colour indices. Since the spectrum of $K$ is bounded~\cite{Patella:2005vx}, the function $\hat{\rho}_\Repr(z)$ can be expanded as a series in powers of $z$:
\begin{equation}
\hat{\rho}_\Repr(z) = \frac{e^{-z m^2}}{d_\Repr V} \langle \Tr e^{zK^2} \rangle =
\frac{e^{-z m^2}}{d_\Repr V} \sum_{n=0}^{\infty} \frac{z^n}{n!} \langle \Tr K^{2n} \rangle \comma
\label{eq:hatrho}
\end{equation}
with an infinite radius of convergence. Thus, $\hat{\rho}(z)$ is an analytical function on the whole complex plane. Using the translational invariance, the trace over the site index can be eliminated:
\begin{equation}
\frac{1}{V} \langle \Tr K^{2n} \rangle = \frac{1}{V} \sum_x \langle \tr K^{2n}(x,x) \rangle = \langle \tr K^{2n}(0,0) \rangle \comma
\label{eq:trk2n}
\end{equation}
where $\tr$ denotes the trace over the colour index alone. In order to obtain the loop expansion, we need to compute explicitly the trace in~(\ref{eq:trk2n}):
\begin{equation}
\tr K^{2n}(0,0) = \sum_{ \left\{ x_i \right\} } \tr K(0,x_1) K(x_1,x_2) \cdots K(x_{2n-1},0) \point
\end{equation}
Since the matrix $K(x,y)$ is different from zero only if $x$ and $y$ are nearest neighbours, the generic term in the sum above is different from zero only if the set ($x_0=0$, $x_1$, $\dots$, $x_{2n-1}$, $x_{2n}=0$) is a closed path with length $2n$, connecting nearest neighbours and departing from $0$. Let $\paths_{2n}$ be the (finite) set of such paths. Moreover if $x$ and $y$ are nearest neighbours, $2 K(x,y) = \eta_{y-x}(x) R[U_{y-x}(x)] $ where $\eta_{-\mu}(x) = -\eta_{\mu}(x)$ and $U_{-\mu}(x) = U_{\mu}(x-\hat{\mu})^\dagger$. Finally we obtain the desired loop expansion:
\begin{eqnarray}
\tr K^{2n}(0,0) &=& \frac{1}{4^n}\sum_{\omega \in \paths_{2n}} \eta_{x_1-x_0}(x_0) \cdots \eta_{x_0-x_{2n}}(x_{2n}) \tr \Repr[U_{x_1-x_0}(x_0) \cdots U_{x_0-x_{2n}}(x_{2n})] = \nonumber \\
&=& \sum_{\omega \in \paths_{2n}} c(\omega) \tr \Repr[U(\omega)] \ , 
\label{eq:trk2n_loops}
\end{eqnarray}
\begin{equation}
c(\omega) = 4^{-n} \eta_{x_1-x_0}(x_0) \cdots \eta_{x_0-x_{2n}}(x_{2n}) 
\end{equation}
and $\omega$ is the closed path $(x_0, x_1, \dots, x_{2n}, x_0)$.

At this stage, we have all the tools we need to prove the orientifold planar equivalence. Putting the formulae~(\ref{eq:hatrho}), (\ref{eq:trk2n}), (\ref{eq:trk2n_loops}) together, we obtain:
\begin{equation}
\hat{\rho}_\Repr(z) = \frac{e^{-z m^2}}{d_\Repr} \sum_{n=0}^{\infty} \frac{z^n}{n!} \sum_{\omega \in \paths_{2n}} c(\omega) \langle \tr \Repr[U(\omega)] \rangle \point
\end{equation}
By taking the large-$N$ limit\footnote{The large-$N$ limit can pass through the infinite sum, since the eigenvalues of $K$ are bounded, uniformly in the number of colours.} and using the Eq.~(\ref{eq:groupequivalence}), it is straightforward to prove that:
\begin{equation}
\lim_{N \rightarrow \infty} \hat{\rho}_\Sym(z) = \lim_{N \rightarrow \infty} \hat{\rho}_\Asym(z) = \lim_{N \rightarrow \infty} \hat{\rho}_\Adj(z) \point
\end{equation}
The equality of the Laplace transforms immediately implies the equality of the EPDs.

The fermionic condensates can be obtained by integrating the Laplace transform of the EPD:
\begin{eqnarray} 
\langle \bar{\psi} \psi \rangle_\Repr &=& \frac{1}{V} \langle \Tr D^{-1} \rangle = \frac{m}{V} \langle \Tr H^{-1} \rangle = \nonumber \\
&=& \frac{m}{V} \int_0^\infty \langle \Tr e^{-tH} \rangle \, dt = m d_\Repr \int_0^\infty \hat{\rho}_\Repr(t) \, dt \point
\label{eq:condensate0}
\end{eqnarray}
The second equality comes observing that $D^{-1}=(m-K)/(m^2-K^2)=m H^{-1} - K H^{-1}$ and the trace of the second term vanishes, since the eigenvalues of $K$ are pairs of opposite purely imaginary numbers. By specialising to the two-index representations and taking the large-$N$ limit\footnote{The large-$N$ limit and the integral can be exchanged because the Laplace transform is controlled by the lowest eigenvalue in the range $t\in[0,\infty)$, i.e. $|\hat{\rho}(t)| \le e^{-tm^2}$.}:
\begin{eqnarray}
&& \lim_{N \rightarrow \infty} \frac{ \langle \bar{\psi} \psi \rangle_{\Sym / \Asym} }{N^2} = 
m \lim_{N \rightarrow \infty} \frac{N(N \pm 1)}{2 N^2} \int_0^\infty \hat{\rho}_{\Sym / \Asym}(t) \, dt = \nonumber \\
&& \qquad = m \lim_{N \rightarrow \infty} \frac{N^2-1}{2 N^2} \int_0^\infty \hat{\rho}_{\Adj}(t) \, dt = 
\lim_{N \rightarrow \infty} \frac{ \langle \bar{\psi} \psi \rangle_{\Adj} }{2N^2} = \lim_{N \rightarrow \infty} \frac{ \langle \lambda \lambda \rangle_{\Adj} }{N^2} \point
\end{eqnarray}

\section{Separating the $1/N$ corrections}
\label{sec:quenched:condensate:corrections}
The fermionic condensate in the (anti)symmetric representation contains odd and even power corrections in $1/N$ to the planar limit. In the quenched theories, some analytical relationships allow to separate the even corrections from the odd ones.

The condensate of fermions in the two-index representations is obtained by computing the integral in Eq.~(\ref{eq:condensate0}), and by using the formulae:
\begin{subequations}
\begin{flalign}
& \tr \Asym[U] = \frac12 \left[ ( \tr U )^2 - \tr (U^2) \right] \comma \\
& \tr \Sym[U] = \frac12 \left[ ( \tr U )^2 + \tr (U^2) \right] \comma \\
& \tr \Adj[U] = | \tr U |^2 -1 \point
\end{flalign}
\end{subequations}
The results for the fermionic condensates are:
\begin{flalign}
\frac{ \langle \bar{\psi} \psi \rangle_{\Sym / \Asym} }{N^2} &=
\left\{ \sum_{n=0}^{\infty} \frac{1}{m^{2n+1}} \sum_{\omega \in \paths_{2n}} c(\omega) \frac{\langle [\tr U(\omega)]^2 \rangle}{2N^2} \right\} \pm \frac1N \left\{ \sum_{n=0}^{\infty} \frac{1}{m^{2n+1}} \sum_{\omega \in \paths_{2n}} c(\omega) \frac{\langle \tr [U(\omega)^2] \rangle}{2N} \right\}
\ , \nonumber \\
\frac{ \langle \lambda \lambda \rangle_{\Adj} }{N^2} &=
\left\{ \sum_{n=0}^{\infty} \frac{1}{m^{2n+1}} \sum_{\omega \in \paths_{2n}} c(\omega) \frac{\langle |\tr U(\omega)|^2 \rangle}{2N^2} \right\} - \frac{1}{2N^2} \left\{ \sum_{n=0}^{\infty} \frac{1}{m^{2n+1}} \sum_{\omega \in \paths_{2n}} c(\omega) \right\}
\point \nonumber \\
&
\end{flalign}
All the expressions in the curly brackets are finite in the large-$N$ limit. Moreover, since they are expectation values of gluonic operators, they contain only powers of $1/N^2$. We will write schematically:
\begin{subequations}
\begin{flalign}
& \frac{ \langle \bar{\psi} \psi \rangle_{\Sym / \Asym} }{N^2} = f\left( \frac{1}{N^2},m \right) \pm \frac{1}{N} g\left( \frac{1}{N^2},m \right) \ , \\
& \frac{ \langle \lambda \lambda \rangle_{\Adj} }{N^2} = \tilde{f}\left( \frac{1}{N^2},m \right) - \frac{1}{2N^2} \langle \bar{\psi} \psi \rangle_{\textrm{free}} \point
\end{flalign}
\end{subequations}
The functions $f$ and $g$ describe respectively the even and odd corrections to the planar limit of the fermionic condensate in the (anti)symmetric representation. They can be computed by inverting the equations above, while $\langle \bar{\psi} \psi \rangle_{\textrm{free}}$ is the condensate of free fermions. The functions $f$ and $\tilde{f}$ are unrelated at generic $N$. However, at large-$N$ the orientifold planar equivalence holds; while at $N=2$ the symmetric and the adjoint are the same representation, and the antisymmetric is the trivial representation. Therefore:
\begin{subequations}
\begin{flalign}
& f(0,m) = \lim_{N\rightarrow \infty} \frac{\langle \bar{\psi} \psi \rangle_{\Sym} + \langle \bar{\psi} \psi \rangle_{\Asym}}{2N^2} = \lim_{N\rightarrow \infty} \frac{\langle \lambda \lambda \rangle_{\Adj}}{N^2} = \tilde{f}(0,m) \comma \\
& f(1/4,m) = \left. \frac{\langle \bar{\psi} \psi \rangle_{\Sym} + \langle \bar{\psi} \psi \rangle_{\Asym}}{2N^2} \right|_{N=2} = \left. \frac{2 \langle \lambda \lambda \rangle_{\Adj} + \langle \bar{\psi} \psi \rangle_{\textrm{free}}}{2N^2} \right|_{N=2} = \tilde{f}(1/4,m) \point \label{eq:feqftilde_at_N2}
\end{flalign}
\end{subequations}

\section{Numerical calculations}
\label{sec:calculation}
We run simulations for $N=2,3,4,6,8$ at constant lattice spacing. The lattice spacing was fixed by simulating at the values of $\beta$ for which the SU($N$) lattice gauge theories have a deconfinement transition for $N_0 = 5$. The values of $\beta$ used in our calculation have been determined in~\cite{Lucini:2003zr}. Requiring a fixed value for a given quantity across the various $N$ is equivalent to fix the normalised 't Hooft coupling. The choice used in this paper has been inspired by a recent work on the large-$N$ meson spectrum~\cite{DelDebbio:2007wk}. In principle another observable (e.g. the string tension) could have been used. Different choices change the size of the $1/N$ corrections, but not the large-$N$ value. Dimensionful units can be reinstated by noting $T_c=(5a)^{-1}$. Using the value $T_c = 270 \ \mathrm{MeV}$, for the lattice spacing $a$ we get $a \simeq 0.145 \ \mathrm{fm}$. We used a $N_0 \times N_s^3$ lattice with the size $N_0 = N_s = 14$ (which care
 responds to about $2.0 \ \mathrm{fm}$). For SU(3), a $16^3 \times 24$ lattice has been used to check for finite size effects; it was found that the difference of the condensates is zero within at most two standard deviations. Since finite size effects are less severe as $N$ increases~\cite{Narayanan:2004cp,Kovtun:2007py}, we can reasonably assume that our results are not affected by finite size artifacts, at least for $N > 3$, i.e. for the values we have used to extract the infinite $N$ behaviour of the condensates.

At each value of $N$ listed above, Monte Carlo ensembles of gauge fields have been generated using the Wilson action. The link variables have been updated using a Cabibbo-Marinari algorithm~\cite{Cabibbo:1982zn}, where each $SU(2)$ subgroup of $SU(N)$ is updated in turn. We have measured the quark condensate for the symmetric, antisymmetric and adjoint representations. For each value of $N$, 100 measurements have been collected, separated by 5 heath-bath and 20 micro-canonical steps alternated in a ratio 1:4. 

The values of the bare mass have been chosen in such a way that all the different regimes of the theories are explored in detail. The algorithm used for the inversion of the Dirac operator is a multi-shift Conjugated Gradient (CG). This enabled us to compute the condensates corresponding to different masses simultaneously. For all the inversions we required a relative precision of $10^{-9}$. We found that the required average number of applications of the Dirac matrix is about $2500$ for masses greater than $1.2 \times 10^{-2}$.

The complete set of bare parameters used in our simulations is summarised in Tab.~\ref{tab:condensate:par}. 

\begin{table}[h]
\begin{center}
\begin{tabular}{|c|c|c|}
\hline
$N$ & $\beta$ & $m$ \\
\hline
\hline
2 & 2.3715 & \multirow{5}{5cm}{0.012, 0.013, 0.014, 0.015, 0.016, 0.017, 0.018, 0.019, 0.02, 0.04, 0.06, 0.08, 0.1, 0.2, 0.4, 0.6, 0.8, 1.0, 2.0, 4.0, 6.0, 8.0}\\
\cline{1-2}
3 & 5.8000 & \\
\cline{1-2}
4 & 10.6370 & \\
\cline{1-2}
6 & 24.5140 & \\
\cline{1-2}
8 & 44.0000 & \\
\hline
\end{tabular}
\caption{Bare parameters used for the computation of the fermionic condensate.}
\label{tab:condensate:par}
\end{center}
\end{table}

The fermionic condensate is computed from formula~(\ref{eq:condensate0}), which we rewrite here for convenience:
\begin{equation}
\langle \bar{\psi} \psi \rangle = \frac{m}{V} \langle \Tr (m^2-K^2)^{-1} \rangle \comma
\label{eq:condensate:condensate}
\end{equation}
where the trace is estimated stochastically with one noisy vector per configuration. In general, these noisy sources are complex. However in the case of the adjoint representation, since the staggered Dirac operator is real, the noisy vector can also be taken real. In the continuum limit, this corresponds to simulating four (since the fermions are staggered) flavours of Dirac fermions in the (anti)symmetric representation or four flavours of Majorana fermions in the adjoint representation.

\begin{figure}
\begin{center}
\includegraphics[angle=270,width=1\textwidth]{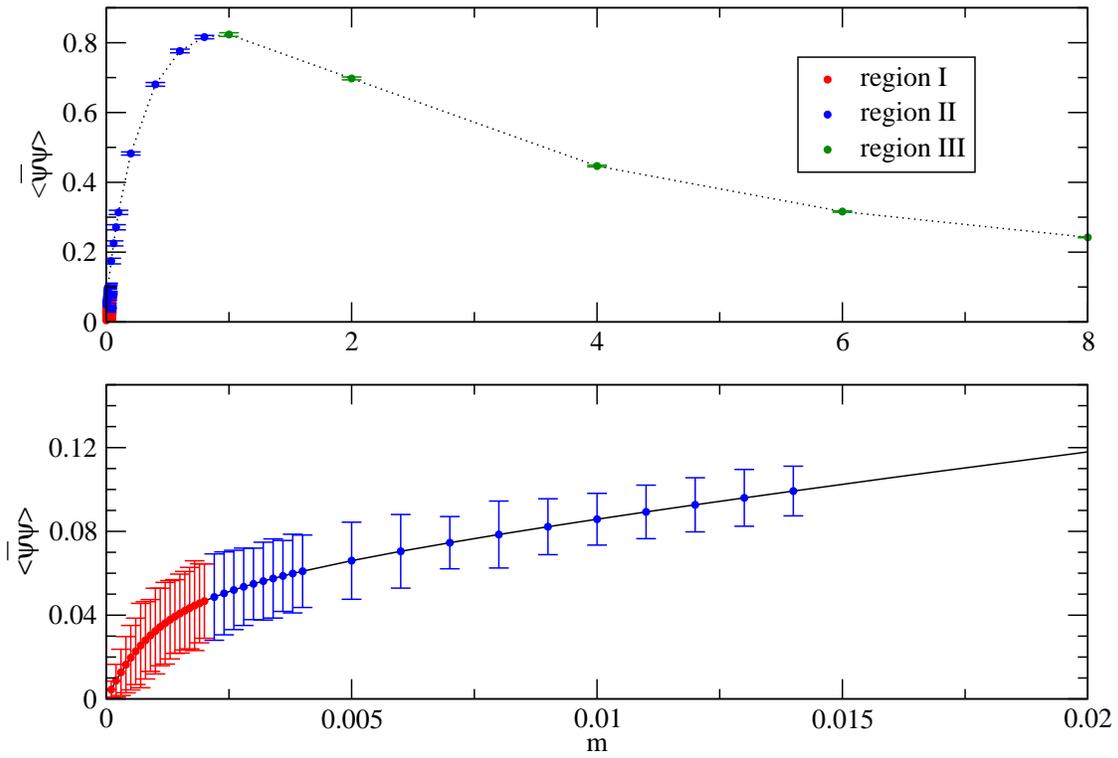}
\caption{\label{fig:condensate:regions} The condensate of staggered fermions in the fundamental representation on a $14^4$ lattice, at $\beta=2.3715$, $N=2$ (in two different scales).}
\end{center}
\end{figure}

As the quark mass changes, the condensate goes through three different regimes (see Fig.~\ref{fig:condensate:regions}). The condensate can be written in terms of the eigenvalues of the massless staggered Dirac operator $D$:
\begin{equation}
\langle \bar{\psi} \psi \rangle = \frac{1}{V} \sum_\alpha \langle \frac{m}{m^2+\lambda_\alpha^2} \rangle \point
\label{eq:condensate:condensate1}
\end{equation}
Since the lattice volume is finite, no chiral symmetry breaking is expected, therefore in the limit of zero mass the condensate is proportional to the mass (\textit{region I}). This is a well-known property of staggered fermions~\cite{Kogut:1983sm,Hands:1990wc}. As the volume increases, the average of the smallest eigenvalue of the operator $K$ approaches zero and the condensate becomes different from zero in the chiral limit. The same behaviour is expected as $N$ increases, since in the planar limit the condensate is independent of the volume~\cite{Narayanan:2004cp,Kovtun:2007py}. However at fixed number of colours, as the mass increases the value of the condensate is less sensitive to the lowest eigenvalues and becomes approximately independent of the volume. Practically, a sudden change in the slope of the condensate occurs, with the condensate still increasing linearly with the mass (\textit{region II}). At some point the condensate starts to decrease (\textit{region III}), and when the mass is much higher than the highest eigenvalue of $K$ the $1/m$ asymptotic behaviour is recovered. Region I is not interesting, since there the condensate is heavily affected by finite volume effects. Thus, the masses have been chosen in such a way to explore the regions II and III.

The condensates for the antisymmetric, symmetric, adjoint representation are reported respectively in Tabs.~\ref{tab:condensate:antisymmetric}, \ref{tab:condensate:symmetric}, \ref{tab:condensate:adjoint}. The data for $N=2$ with the antisymmetric representation are obtained by computing exactly (no random noise) the condensate for the free fermion, and the result were conventionally truncated to the third digit.

Using the data at finite $N$ we have computed the functions $f$, $g$, $\tilde{f}$ defined in Sect.~\ref{sec:quenched:condensate:corrections}, which we report here for convenience:
\begin{subequations}
\begin{flalign}
& \frac{1}{N^2} \langle \bar{\psi} \psi \rangle_{\Sym / \Asym} = f\left( \frac{1}{N^2},m \right) \pm \frac{1}{N} g\left( \frac{1}{N^2},m \right) \comma \\
& \frac{1}{N^2} \langle \lambda \lambda \rangle_{\Adj} = \tilde{f}\left( \frac{1}{N^2},m \right) - \frac{1}{2N^2} \langle \bar{\psi} \psi \rangle_{\textrm{free}} \point
\end{flalign}
\end{subequations}

\begin{figure}
\begin{center}
\includegraphics[width=0.85\textwidth]{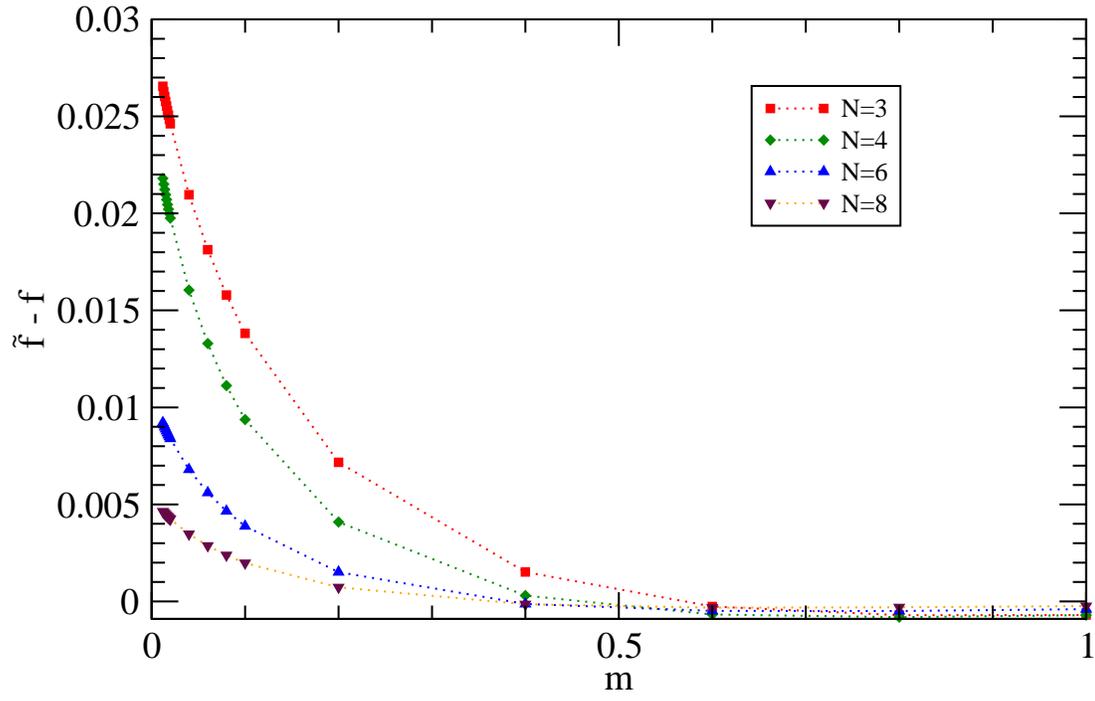}
\caption{\label{fig:condensate:fftildediff} The difference between the functions $\tilde f$ and $f$ as a function of the mass. The difference at $N=2$ is not plotted, since it is zero (into the statistical errors) accordingly with the Eq.~\eqref{eq:feqftilde_at_N2}.}
\end{center}
\end{figure}

Planar equivalence means that $f$ and $\tilde{f}$ are identical in the large-$N$ limit. The functions $f$, $g$ and $\tilde{f}$ can be obtained by fitting the numerical data at finite $N$ with the predicted large $N$-behaviour truncated to a reasonable order in $N$, in analogy to the analysis performed in the literature for glueball~\cite{Lucini:2004my} or meson masses~\cite{DelDebbio:2007wk}. Hence, for the extrapolation we could attempt a fit of the form
\beq
\label{eq:asymptotic}
f\left( \frac{1}{N^2},m \right) = f\left( 0,m \right) +\frac{b(m)}{N^2} \ ,
\eeq
and analogously for $\tilde{f}$ and $g$. However, assuming that such a fit works, the precision of the data poses a potential problem. The condensate is measured with a typical precision of $10^{-3}$ (and below). When performing an extrapolation to large $N$ with this level of precision (which is about one order of magnitude smaller than the precision typically obtained for glueball masses or meson masses), in order to get a reliable large-$N$ result, it is crucial to estimate the error associated with the truncation of the large-$N$ series, which often turns out to be also of order $10^{-3}$. Hence, when extrapolating to $N \to \infty$, we must allow the possibility to add an extra term in the series, to check the magnitude of the truncation error. This proved to be feasible for $\tilde{f}$, for which we obtain a good fit for $N \ge 3$, but not for $f$ and $g$, for which the value $N = 3$ does not provide a good fit. For this reason, in order to prove numerically planar equivalence, instead of comparing the infinite $N$ values of $f$ and $\tilde{f}$ directly, we extrapolated the difference between those two functions:
\beq
\label{eq:diff_largeN}
f\left( \frac{1}{N^2},m \right) - \tilde{f}\left( \frac{1}{N^2},m \right) = \frac{c_1(m)}{N^2} + \frac{c_2(m)}{N^4} \ .
\eeq
The difference $f\left( \frac{1}{N^2},m \right) - \tilde{f}\left( \frac{1}{N^2},m \right)$ is plotted in Fig.~\ref{fig:condensate:fftildediff}. Results for the fits at the various values of $m$ for $N \ge 4$ (the smallest value of $N$ for which we obtain a reasonable fit) are reported in Tab.~\ref{tab:fdiff}. The ansatz works very well, giving a reduced $\chi^2$ always less than one. This result supports the validity of orientifold planar equivalence for the condensates. 

The large $N$ limit of $\tilde{f}$ can be determined fitting a functional form similar to that given in Eq.~(\ref{eq:asymptotic}). This function can be fitted for $N \ge 3$ (although a fairly better reduced $\chi^2$ is obtained excluding the point $N = 3$); this allows us to control the truncation error by inserting a term $1/N^4$ in the ansatz~(\ref{eq:asymptotic}). The truncation error can be accounted for by increasing the errors on the fitted parameters (reported in Tab.~\ref{tab:condensate:ftilde}) by roughly 30\%. Once $\tilde{f}$ has been determined, we can reconstruct $f(1/N^2,m)$ inserting the results of Tab.~\ref{tab:fdiff} in Eq.~(\ref{eq:diff_largeN}). This provide a more robust estimate than extracting $f(0,m)$ from the data of the condensates.

\begin{figure}
\begin{center}
\includegraphics[width=0.85\textwidth]{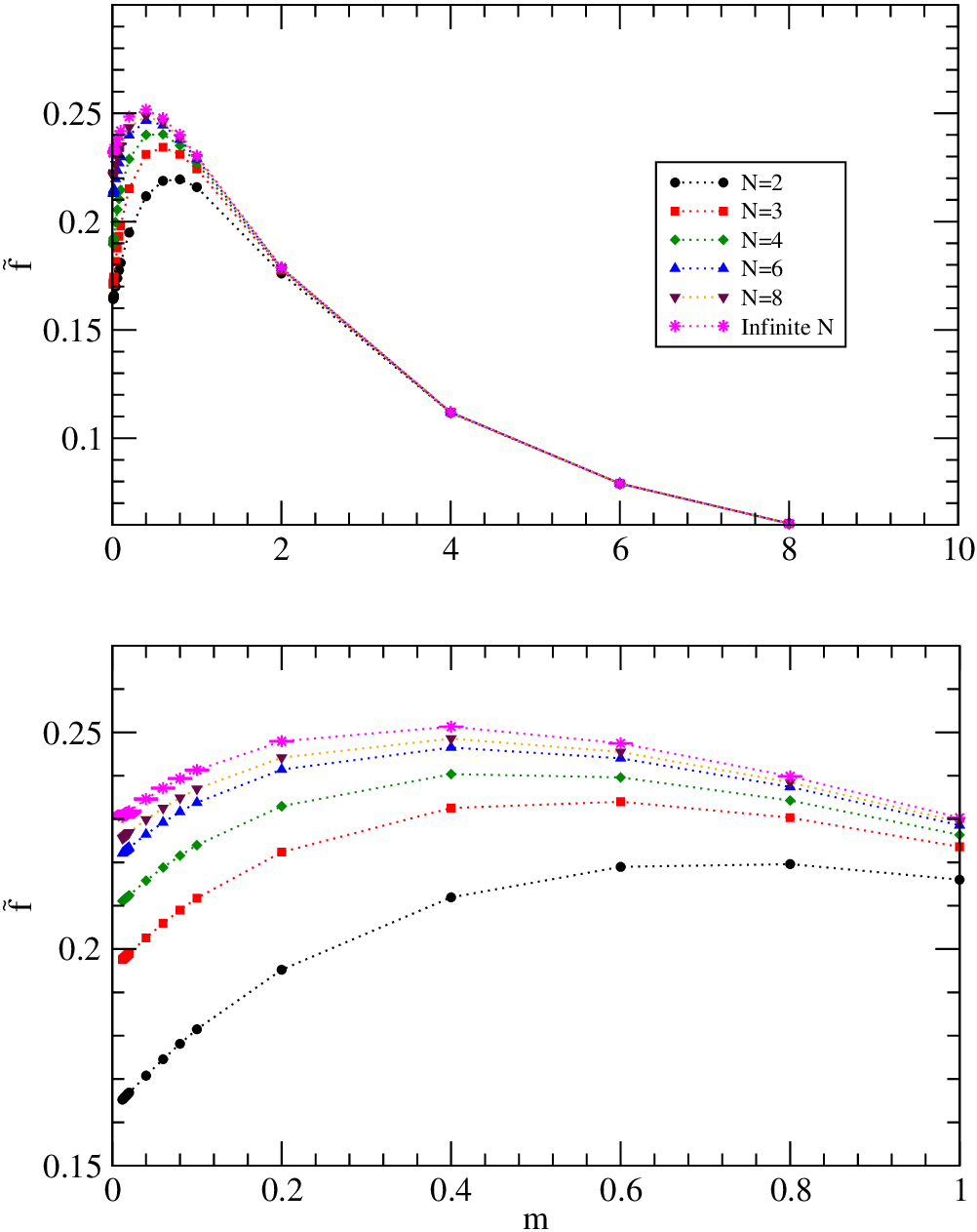}
\caption{\label{fig:condensate:functs_ftilde} The function $\tilde{f}$. The bottom figure is an enlargement of the region $m \in (0, 1)$.}
\end{center}
\end{figure}

\begin{figure}
\begin{center}
\includegraphics[width=0.85\textwidth]{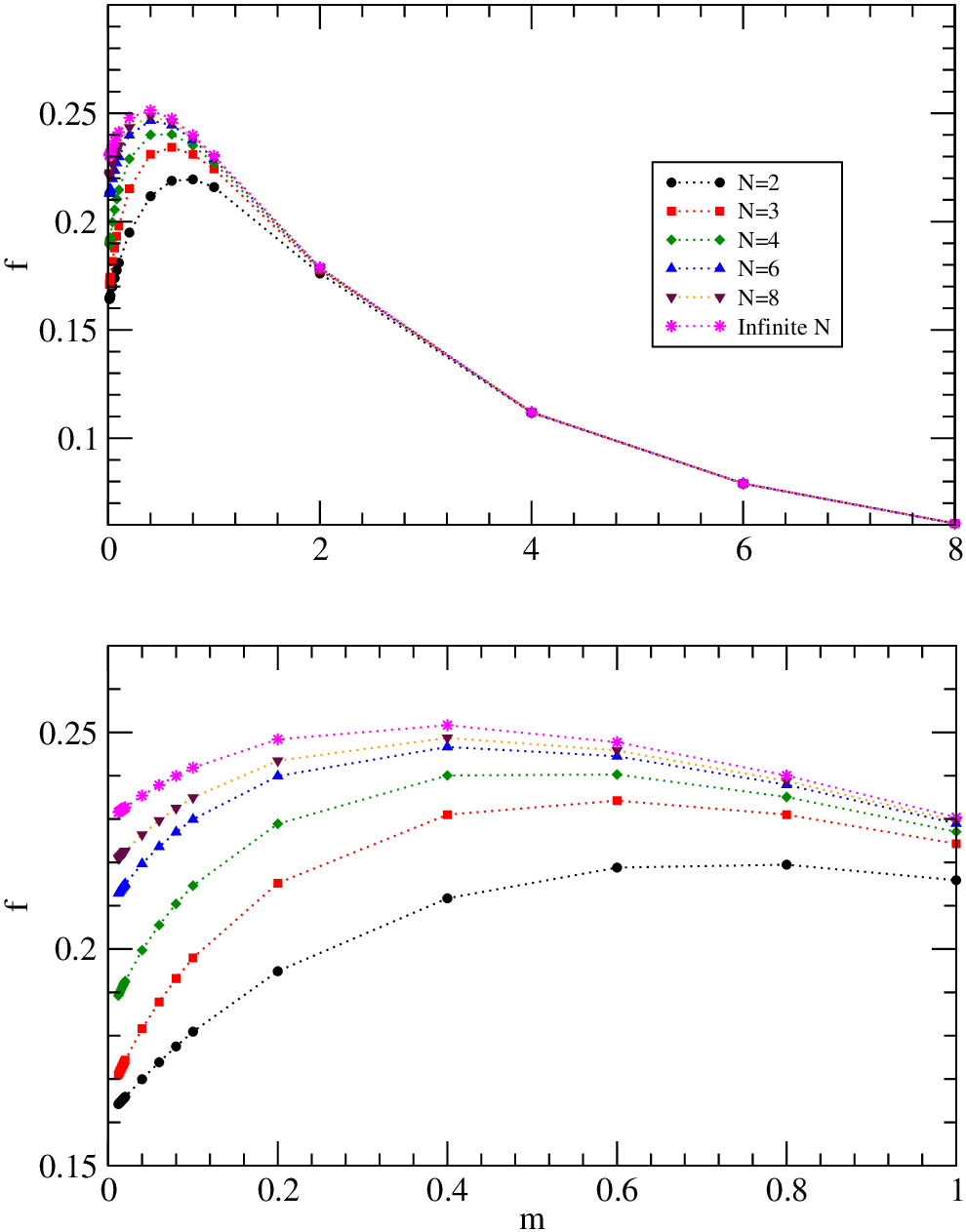}
\caption{\label{fig:condensate:functs_f} The function $f$. The bottom figure is an enlargement of the region $m \in (0, 1)$.}
\end{center}
\end{figure}

\begin{figure}
\begin{center}
\includegraphics[width=0.85\textwidth]{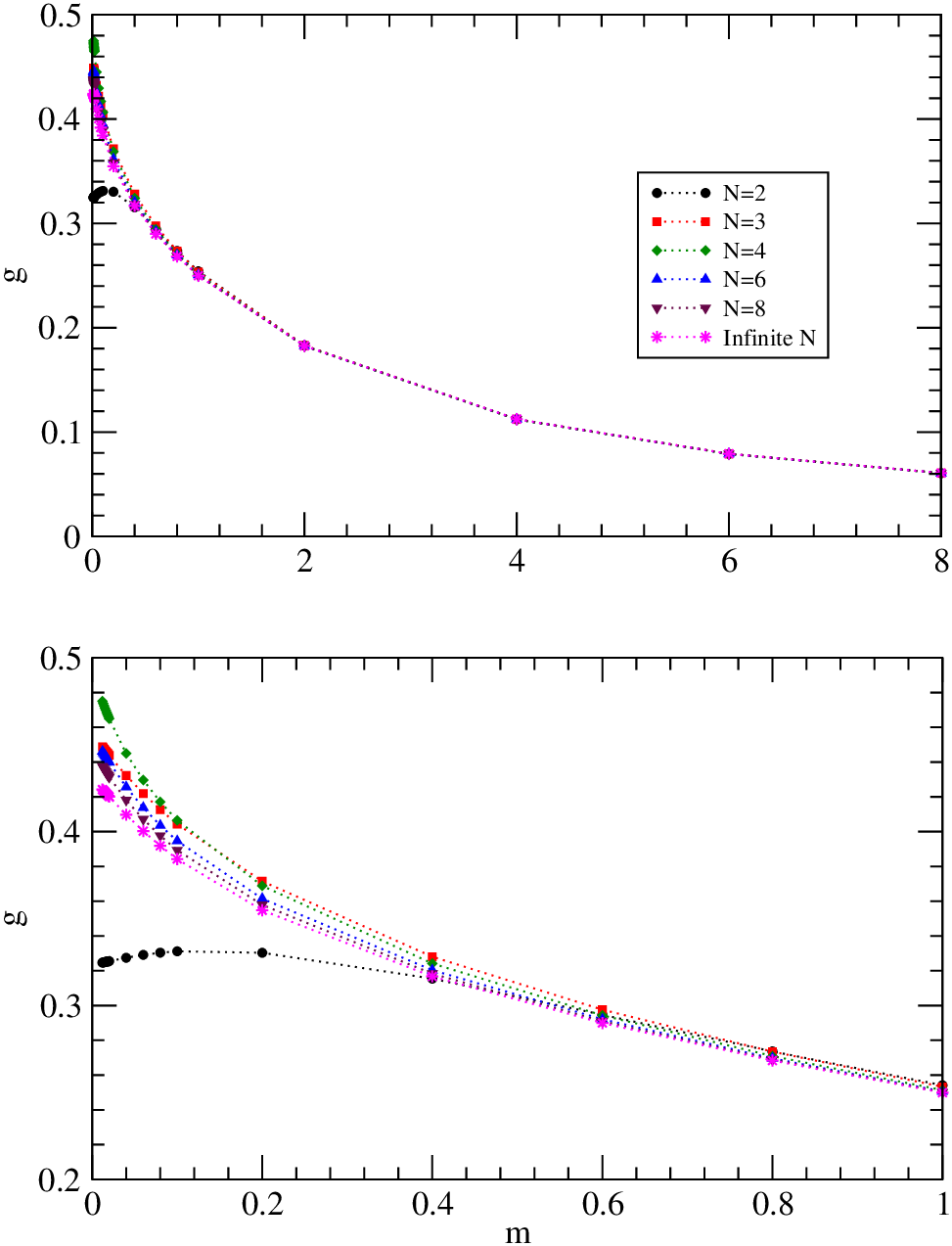}
\caption{\label{fig:condensate:functs_g} The function $g$. The bottom figure is an enlargement of the region $m \in (0, 1)$.}
\end{center}
\end{figure}

The function $g$ turns out to be non-monotonic in $N$: it is increasing for $N=2,3$, attains a maximum for $N=4$ and then decreases for lager $N$. For this reason, the fit of $g$ would present in principle the same problem as the fit of $f$, since also in this case reasonable $\chi^2$ can be obtained only excluding data for $N < 4$. However, since $g$ is related to subleading effects, it is less crucial to have a precise estimate of the error. Hence, for $g$ we perform a simple fit with a ${\cal O}(1/N^2)$ correction. The results of the fits are reported in Tab.~\ref{tab:condensate:g}. The functions $f$, $g$ and $\tilde{f}$ for $N=2,3,4,6,8,\infty$ are displayed in the plots~\ref{fig:condensate:functs_ftilde},~\ref{fig:condensate:functs_f} and~\ref{fig:condensate:functs_g}.


The chiral limit of the condensate is obtained by extrapolating to zero mass each fitted coefficient in the parameterisation of $\tilde{f}$, $\tilde{f} - f$ and $g$. In the interval of masses $(0.012, 0.04)$  we fit those coefficients with a linear function of the mass $h(m) = d_0 + d_1m$. The errors are estimated by checking the stability of the fit when a $d_2 m^2$ term is added. In the chiral limit, our results are summarized by the following parameterisations for the condensates: 
\begin{subequations}
\begin{flalign}
& \frac{1}{N^2} \langle \bar{\psi} \psi \rangle_{\Sym}(m=0) = 0.22881(36) +\frac{0.430461(14)}{N} -\frac{0.649(22)}{N^2} +\frac{0.888(95)}{N^3} +\dots \comma \label{cond-fit1} \\
& \frac{1}{N^2} \langle \bar{\psi} \psi \rangle_{\Asym}(m=0) = 0.22881(36) -\frac{0.430461(14)}{N} -\frac{0.649(22)}{N^2} -\frac{0.888(95)}{N^3} +\dots \comma \label{cond-fit2} \\
& \frac{1}{N^2} \langle \lambda \lambda \rangle_{\Adj}(m=0) = 0.22881(36) -\frac{0.3152(66)}{N^2} +\dots \point
\label{cond-fit3}
\end{flalign}
\end{subequations}

At this point, some comments are in order. The parametrizations above describe well all the data for $N \ge 4$ (the fits were all obtained in this range). It is interesting to compare the extrapolations at $N=3$ and $N=2$ obtained from the truncated series with the data from the simulations. The difference between the extrapolated value and the measured one is the size of the subleading terms. In order to avoid the technicalities related to the chiral extrapolation, we can compare these values at $m=0.012$ for instance. The results are summarized in Tab.~\ref{tab:sistematics}. It is impressive how well a single $1/N^2$ correction describes the condensate in the adjoint representation up to $N=2$. The $1/N^3$ correction describes quite well the condensate for the symmetric representation at $N=3$. Instead, the extrapolation for the antisymmetric representation works very badly. This can be ascribed to the fact that for $N=2,3$ the leading $1/N$ correction is large and opposite in sign to the infinite $N$ value. As a consequence, for these values of $N$ the contribution of higher order corrections are enhanced.

\begin{table}[h]
\begin{center}
\begin{tabular}{|c|c|c|c|c|c|c|}
\hline
                   & \multicolumn{3}{|c|}{$N=3$}             & \multicolumn{3}{|c|}{$N=2$}                 \\
\hline
repr.              & extrap.     & simul.      & rel. error  & extrap.      & simulat.    & rel. error     \\
\hline
\hline
$\Adj$             & 1.76        & 1.7743(48)  & $0.8\%$     & 0.609        & 0.6572(45)  & $8\%$          \\
\hline
$\Asym$            & $<0$        & 0.1927(15)  & -           & $<0$         & 0.00746     & -              \\
\hline
$\Sym$             & 3.01        & 2.8854(75)  & $4\%$       & 1.56         & 1.3062(69)  & $16\%$         \\
\hline
\end{tabular}
\caption{In this Table, the values of the condensate at $m=0.012$ are summarised, as obtained by taking $N=3$ or $N=2$ in the truncated series expansions (\textit{extrap.}) or by the numerical simulations (\textit{simul.}). The \textit{relative error} is defined as $|\textit{extrap.}-\textit{simul.}|/\textit{extrap}$.}
\label{tab:sistematics}
\end{center}
\end{table}

\newpage 

It was proposed \cite{Armoni:2003yv} (see also \cite{Armoni:2005wt}) that the ratio between the chiral-limit quark condensate in the antisymmetric and adjoint representations should be well approximated by the ratio between the Dynkin index of the two representations, namely that 
\beq
\frac{ \langle \bar{\psi} \psi \rangle_{\Asym} }{ \langle \lambda \lambda \rangle_{\Adj} } \sim \frac{N-2}{N} \, . \label{ratio}
\eeq 
Let us compare the above assertion \eqref{ratio} to our fit \eqref{cond-fit1},\eqref{cond-fit2},\eqref{cond-fit3}. The ratio of the quark condensate in the antisymmetric over the adjoint can be computed from the Eqs.~\eqref{cond-fit2} and~\eqref{cond-fit3} and expressed (by truncating the series expansion) as follows
\beq
\frac{ \langle \bar{\psi} \psi \rangle_{\Asym} }{ \langle \lambda \lambda \rangle_{\Adj} } = \frac{N-2}{N} \left( 1 + \frac{0.1189(2)}{N} - \frac{1.22(7)}{N^2}  + \dots\right) \, . \label{ratio2}
\eeq 
Notice that the overall factor $\frac{N-2}{N}$ captures indeed the first $1/N$ correction, as proved by the small coefficient $0.1189$. However, the sub-leading $1/N^2$ correction cannot be neglected for low values of $N$ (and in particular for $SU(3)$) due to the large numerical coefficient $1.22$. It would be interesting to study how the full unquenched theory would modify the result \eqref{ratio2}.

\section{Conclusions}
\label{sec:conclusions}
In this paper, after having shown analytically the equality between fermionic condensates in the (anti)symmetric and the adjoint representation for quenched SU($N$) gauge theory in the infinite $N$ limit, we have checked its validity on the lattice. We found a satisfying agreement, hence our data support the proposed equivalence. 

One could object that the analytical proof makes the numerical check unnecessary. As a matter of fact, there are at least two motivations to perform a lattice simulation. First, the analytical argument assumes charge conjugation. Although reasonable and not contradicting any evidence we have from QCD-like theories, this assumption mandates a first principle calculation. Then, the proof holds in the limit $N = \infty$; it is still an interesting question to understand what happens at finite $N$. Once again, we remark that although the analytical proof was given on the lattice, it is easy to reproduce it in regularisation schemes more suitable for investigating the theory directly in the continuum. On the other hand, formulating the proof on the lattice enabled us to show explicitly that the equivalence holds at any fixed lattice spacing (hence, also in the continuum limit). This fact was exploited by performing the numerical simulations at only one value of the lattice spacing, which allowed us to circumvent technical problems like the renormalisation of the condensate in the continuum limit.

For the chiral condensate, numerical studies of the approach to the large-$N$ limit are complicated by the presence of a ${\cal O}(1/N)$ correction in the (anti)symmetric representation, which delay the onset of the limit to larger values of $N$ with respect to other pure gauge observables (see e.g.~\cite{Lucini:2003zr,DelDebbio:2002xa,Del Debbio:2006df,Lucini:2004my,Lucini:2004yh} for numerical studies of pure gauge observables in the large-$N$ limit). In particular, if we compare with~\cite{DelDebbio:2007wk}, where the quenched meson spectrum was investigated at the same values of $\beta$, we notice that in this work it has proven to be necessary to go to $N=8$, while for the meson spectrum $N = 6$ was sufficient. Moreover, to achieve a precise result, it was crucial to separate the odd contributions in powers of $1/N$ from the even ones. This was obtained by using at the same time information from the symmetric and the antisymmetric representation. Using this method, we were able to extrapolate to infinite $N$ and to prove numerically the equality between condensates of Dirac fermions in the (anti)symmetric representation and of Majorana fermions in the adjoint representation by extrapolating to $N = \infty$ their difference. We expect the techniques we have used in this work to be relevant for a numerical study of planar equivalence in theories with dynamical fermions.

\section*{Acknowledgements}
We thank L. Del Debbio and S. Hands for stimulating discussions, and M. Shifman and G. Veneziano for comments on the manuscript. Simulations have been performed on a computer cluster partially funded by the Royal Society and STFC. A.A. is supported by the PPARC advanced fellowship award. B.L. is supported by a Royal Society fellowship. A.P. is supported by an STFC special project grant and by the ``Fondazione Angelo Della Riccia''. The work of C.P. has been supported by contract No. DE-AC02-98CH10886 with the U.S. Department of Energy.

\newpage

\section*{Appendix}
To enable the reader to perform an independent analysis, in this Appendix we
collect tables of the numerical values for the condensates at various $N$ and
fit parameters, as discussed in Sect.~\ref{sec:calculation}, to which we refer
for details.~\\
\vspace{4cm}

\begin{table}[h]
\begin{center}
\begin{tabular}{|c||c|c|c|c|c|}
\hline
$m$ & $N=2$ & $N=3$ & $N=4$ & $N=6$ & $N=8$ \\
\hline
0.012 & 0.00746 & 0.1927(15)    & 1.1279(68)   & 4.985(12)  &  10.639(15) \\
0.013 & 0.00808 & 0.1985(15)    & 1.1404(65)   & 5.000(12)  &  10.660(15) \\
0.014 & 0.00870 & 0.2043(15)    & 1.1526(62)   & 5.015(11)  &  10.680(15) \\
0.015 & 0.00932 & 0.2100(14)    & 1.1647(62)   & 5.029(10)  &  10.700(14) \\
0.016 & 0.00995 & 0.2156(14)    & 1.1765(58)   & 5.044(10)  &  10.720(14) \\
0.017 & 0.0106  & 0.2212(14)    & 1.1882(56)   & 5.0583(98) &  10.740(14) \\
0.018 & 0.0112  & 0.2267(13)    & 1.1997(54)   & 5.0725(94) &  10.760(13) \\
0.019 & 0.0118  & 0.2322(13)    & 1.2110(52)   & 5.0866(92) &  10.778(13) \\
0.02  & 0.0124  & 0.2377(13)    & 1.2221(51)   & 5.1006(89) &  10.798(12) \\
0.04  & 0.0248  & 0.3377(10)    & 1.4156(34)   & 5.3541(62) &  11.145(10) \\
0.06  & 0.0371  & 0.42442(92)   & 1.5610(27)   & 5.5673(53) &  11.4417(88)\\
0.08  & 0.0493  & 0.50078(84)   & 1.6983(23)   & 5.7502(47) &  11.6990(79)\\
0.1   & 0.0614  & 0.56876(78)   & 1.8079(20)   & 5.9093(43) &  11.9246(71)\\
0.2   & 0.119   & 0.82224(61)   & 2.1868(15)   & 6.4678(30) &  12.7189(53)\\
0.4   & 0.216   & 1.09482(50)   & 2.5441(12)   & 6.9572(20) &  13.3715(39)\\
0.6   & 0.286   & 1.21556(45)   & 2.66721(97)  & 7.0497(15) &  13.4057(32)\\
0.8   & 0.331   & 1.25893(42)   & 2.67776(86)  & 6.9465(13) &  13.1300(28)\\
1.0   & 0.355   & 1.25887(41)   & 2.62768(77)  & 6.7391(11) &  12.6903(24)\\
2.0   & 0.338   & 1.05022(34)   & 2.12176(55)  & 5.33279(77)&  9.9744(16) \\
4.0   & 0.223   & 0.66995(23)   & 1.34270(35)  & 3.35753(48)&  6.26867(93)\\
6.0   & 0.158   & 0.47401(16)   & 0.94922(26)  & 2.37235(34)&  4.42842(65)\\
8.0   & 0.121   & 0.36359(12)   & 0.72797(20)  & 1.81921(26)&  3.39574(50)\\
\hline
\end{tabular}
\caption{Results for the condensate of antisymmetric Dirac fermions.}
\label{tab:condensate:antisymmetric}
\end{center}
\end{table}

\begin{table}
\begin{center}
\begin{tabular}{|c||c|c|c|c|c|}
\hline
$m$ & $N=2$ & $N=3$ & $N=4$ & $N=6$ & $N=8$ \\
\hline
0.012 & 1.3062(69) & 2.8854(75) & 4.9281(83) & 10.339(12)   & 17.641(27) \\
0.013 & 1.3072(66) & 2.8875(71) & 4.9298(80) & 10.345(11)   & 17.649(27) \\
0.014 & 1.3083(63) & 2.8896(69) & 4.9316(77) & 10.350(11)   & 17.656(26) \\
0.015 & 1.3093(61) & 2.8915(67) & 4.9332(74) & 10.355(11)   & 17.664(25) \\
0.016 & 1.3104(59) & 2.8935(65) & 4.9349(72) & 10.361(11)   & 17.671(24) \\
0.017 & 1.3114(57) & 2.8954(63) & 4.9366(70) & 10.366(10)   & 17.678(24) \\
0.018 & 1.3124(55) & 2.8972(61) & 4.9383(68) & 10.370(10)   & 17.686(23) \\
0.019 & 1.3135(54) & 2.8990(60) & 4.9400(67) & 10.3752(99)  & 17.693(23) \\
0.02  & 1.3145(52) & 2.9008(59) & 4.9417(65) & 10.3799(97)  & 17.700(22) \\
0.04  & 1.3348(38) & 2.9314(44) & 4.9755(49) & 10.4625(75)  & 17.836(15) \\
0.06  & 1.3537(31) & 2.9559(36) & 5.0071(41) & 10.5324(64)  & 17.956(12) \\
0.08  & 1.3708(28) & 2.9766(32) & 5.0353(37) & 10.5928(59)  & 18.0618(98)\\
0.1   & 1.3861(25) & 2.9944(29) & 5.0599(34) & 10.6449(54)  & 18.1537(83)\\
0.2   & 1.4401(18) & 3.0511(23) & 5.1374(25) & 10.8055(40)  & 18.4450(52)\\
0.4   & 1.4776(13) & 3.0635(17) & 5.1378(17) & 10.8010(28)  & 18.4649(37)\\
0.6   & 1.4645(11) & 3.0011(14) & 5.0229(13) & 10.5544(23)  & 18.0601(32)\\
0.8   & 1.42505(93)& 2.8995(12) & 4.8454(12) & 10.1819(20)  & 17.4327(30)\\
1.0   & 1.37177(85)& 2.7782(10) & 4.6384(10) & 9.74604(17)  & 16.6930(28)\\
2.0   & 1.07038(62)& 2.14857(65)& 3.58079(76)& 7.52300(12)  & 12.8953(21)\\
4.0   & 0.67148(39)& 1.34442(38)& 2.23947(48)& 4.704814(70) & 8.0665(13) \\
6.0   & 0.47428(28)& 0.94930(27)& 1.58119(34)& 3.321850(49) & 5.69551(91)\\
8.0   & 0.36367(21)& 0.72786(20)& 1.21234(26)& 2.546928(37) & 4.36689(70)\\
\hline
\end{tabular}
\caption{Results for the condensate of symmetric Dirac fermions.}
\label{tab:condensate:symmetric}
\end{center}
\end{table}

\begin{table}
\begin{center}
\begin{tabular}{|c||c|c|c|c|c|}
\hline
$m$ & $N=2$ & $N=3$ & $N=4$ & $N=6$ & $N=8$ \\
\hline
0.012 &  0.6572(45)  & 1.7743(48) & 3.3731(88)  & 7.990(13)   & 14.433(20)   \\
0.013 &  0.6577(43)  & 1.7755(47) & 3.3752(84)  & 7.996(13)   & 14.443(19)   \\
0.014 &  0.6582(42)  & 1.7768(46) & 3.3774(82)  & 8.002(13)   & 14.454(18)   \\
0.015 &  0.6588(41)  & 1.7780(45) & 3.3797(80)  & 8.008(12)   & 14.464(17)   \\
0.016 &  0.6593(40)  & 1.7793(44) & 3.3821(77)  & 8.014(12)   & 14.474(16)   \\
0.017 &  0.6597(39)  & 1.7806(43) & 3.3845(76)  & 8.019(12)   & 14.484(16)   \\
0.018 &  0.6602(38)  & 1.7820(43) & 3.3869(73)  & 8.025(12)   & 14.494(15)   \\
0.019 &  0.6607(37)  & 1.7833(42) & 3.3893(72)  & 8.031(12)   & 14.504(15)   \\
0.02  &  0.6612(36)  & 1.7846(41) & 3.3918(70)  & 8.036(12)   & 14.514(14)   \\
0.04  &  0.6706(23)  & 1.8108(32) & 3.4399(50)  & 8.1406(97)  & 14.7005(87)  \\
0.06  &  0.6796(17)  & 1.8348(27) & 3.4826(40)  & 8.2330(84)  & 14.8641(68)  \\
0.08  &  0.6878(13)  & 1.8562(24) & 3.5201(34)  & 8.3144(73)  & 15.0081(58)  \\
0.1   &  0.6952(11)  & 1.8752(22) & 3.5531(30)  & 8.3860(65)  & 15.1350(52)  \\
0.2   &  0.72145(67) & 1.9418(17) & 3.6681(18)  & 8.6317(39)  & 15.5689(36)  \\
0.4   &  0.73971(51) & 1.9849(12) & 3.7379(12)  & 8.7665(24)  & 15.8009(23)  \\
0.6   &  0.73293(45) & 1.9630(10) & 3.69089(99) & 8.6415(20)  & 15.5693(18)  \\
0.8   &  0.71310(41) & 1.90758(91)& 3.58327(87) & 8.3812(18)  & 15.0965(15)  \\
1.0   &  0.68638(38) & 1.83459(84)& 3.44408(80) & 8.0505(17)  & 14.4986(14)  \\
2.0   &  0.53553(27) & 1.42893(62)& 2.67948(61) & 6.2560(13)  & 11.26308(97) \\
4.0   &  0.33594(17) & 0.89587(39)& 1.67940(40) & 3.91990(83) & 7.05645(63)  \\
6.0   &  0.23728(13) & 0.63271(28)& 1.18603(28) & 2.76827(59) & 4.98324(45)  \\
8.0   &  0.181942(96)& 0.48513(21)& 0.90940(21) & 2.12259(45) & 3.82091(35)  \\
\hline
\end{tabular}
\caption{Results for the condensate of adjoint Majorana fermions.}
\label{tab:condensate:adjoint}
\end{center}
\end{table}

\begin{table}
\begin{center}
\begin{tabular}{|c||c|c|c|}
\hline
\rule[-1ex]{0pt}{3.5ex} $m$ & $c_1$ & $c_2$ & $\chi^2/\textrm{dof}$ \\
\hline
0.012 &  0.302(36)   & 0.77(70)   & 0.29            \\
0.013 &  0.298(35)   & 0.75(67)   & 0.31            \\
0.014 &  0.295(34)   & 0.73(65)   & 0.32            \\
0.015 &  0.291(33)   & 0.72(63)   & 0.34            \\
0.016 &  0.288(32)   & 0.71(61)   & 0.35            \\
0.017 &  0.285(31)   & 0.70(60)   & 0.35            \\
0.018 &  0.281(30)   & 0.69(58)   & 0.36            \\
0.019 &  0.278(30)   & 0.68(57)   & 0.36            \\
0.02  &  0.275(29)   & 0.67(55)   & 0.37            \\
0.04  &  0.222(21)   & 0.57(40)   & 0.34            \\
0.06  &  0.182(18)   & 0.49(33)   & 0.30            \\
0.08  &  0.150(15)   & 0.45(28)   & 0.26            \\
0.1   &  0.124(14)   & 0.42(25)   & 0.22            \\
0.2   &  0.0424(91)  & 0.37(17)   & 0.10            \\
0.4   &  -0.0132(61) & 0.29(12)   & 0.03            \\
0.6   &  -0.0235(50) & 0.204(94)  & $< 10^{-2}$     \\
0.8   &  -0.0216(44) & 0.138(82)  & $< 10^{-2}$     \\
1.0   &  -0.0169(40) & 0.089(75)  & $< 10^{-2}$     \\
2.0   &  -0.0028(28) & 0.001(54)  & $< 10^{-2}$     \\
4.0   &  0.0004(18)  & -0.012(34) & $< 10^{-2}$     \\
6.0   &  0.0004(13)  & -0.010(24) & $< 10^{-2}$     \\
8.0   &  0.00035(97) & -0.008(19) & $< 10^{-2}$     \\
\hline
\end{tabular}
\caption{Results for the fits of the difference $\tilde f(0,m) - f(0,m)$. $c_1$ and $c_2$ are respectively the coefficient of the ${\cal O}(1/N^2)$ and of the ${\cal O}(1/N^4)$ terms. The reduced $\chi^2$ is also listed.}
\label{tab:fdiff}
\end{center}
\end{table}

\begin{table}
\begin{center}
\begin{tabular}{|c||c|c|c|}
\hline
\rule[-1ex]{0pt}{3.5ex} $m$ & $a(\tilde{f})$ & $b(\tilde{f})$ & $\chi^2/\textrm{dof}$ \\
\hline
0.012 & 0.23050(22)   & -0.3097(72)    & 0.29  \\
0.013 & 0.23067(23)   & -0.3099(75)    & 0.34  \\
0.014 & 0.23083(24)   & -0.3099(77)    & 0.38  \\
0.015 & 0.23099(24)   & -0.3099(78)    & 0.42  \\
0.016 & 0.23115(24)   & -0.3098(78)    & 0.45  \\
0.017 & 0.23131(24)   & -0.3096(78)    & 0.48  \\
0.018 & 0.23147(24)   & -0.3094(78)    & 0.50  \\
0.019 & 0.23162(24)   & -0.3091(77)    & 0.52  \\
0.02  & 0.23177(23)   & -0.3088(76)    & 0.53  \\
0.04  & 0.23462(17)   & -0.3005(59)    & 0.65  \\
0.06  & 0.23714(15)   & -0.2921(51)    & 0.79  \\
0.08  & 0.23935(13)   & -0.2841(47)    & 0.92  \\
0.1   & 0.24131(13)   & -0.2763(43)    & 1.04  \\
0.2   & 0.247952(92)  & -0.2392(31)    & 1.26  \\
0.4   & 0.251322(49)  & -0.1750(16)    & 0.82  \\
0.6   & 0.247472(25)  & -0.12559(84)   & 0.34  \\
0.8   & 0.239862(13)  & -0.08914(45)   & 0.12  \\
1.0   & 0.2303030(10) & -0.06306(24)   & 0.05  \\
2.0   & 0.1788250(10) & -0.012645(32)  & $< 10^{-2}$\\
4.0   & 0.1120220(11) & -0.001669(41)  & $< 10^{-2}$ \\
6.0   & 0.0791087(11) & -0.000726(41)  & 0.01  \\
8.0   & 0.0606567(10) & -0.000486(36)  & 0.01  \\
\hline
\end{tabular}
\caption{Results for the fit of the function $\tilde{f}$ for $N \ge 4$. $a$ is the leading term in the large-$N$ limit, while $b$ is the coefficient of the ${\cal O}(1/N^2)$ correction. The reduced $\chi^2$ is also listed.}
\label{tab:condensate:ftilde}
\end{center}
\end{table}

\begin{table}
\begin{center}
\begin{tabular}{|c||c|c|c|}
\hline
\rule[-1ex]{0pt}{3.5ex} $m$ & $a(g)$ & $b(g)$ & $\chi^2/\textrm{dof}$ \\
\hline
0.012 & 0.4242(11)    &	0.811(25)     &	  0.17          \\
0.013 & 0.42371(95)   &	0.798(21)     &   0.13          \\
0.014 & 0.42321(80)   &	0.785(18)     &   0.10          \\
0.015 & 0.42270(66)   &	0.773(15)     &   0.07          \\
0.016 & 0.42219(53)   &	0.761(12)     &   0.05          \\
0.017 & 0.42167(42)   &	0.7494(95)    &   0.03          \\
0.018 & 0.42115(33)   &	0.7383(73)    &   0.02          \\
0.019 & 0.42062(24)   &	0.7276(53)    &   0.01          \\
0.02  & 0.42010(16)   &	0.7173(35)    &   $ < 10^{-2}$  \\
0.04  & 0.40975(54)   &	0.564(12)     &   0.12          \\
0.06  & 0.40030(73)   &	0.471(16)     &   0.30          \\
0.08  & 0.39183(79)   &	0.406(17)     &   0.46          \\
0.1   & 0.38421(80)   &	0.358(18)     &   0.60          \\
0.2   & 0.35474(71)   &	0.227(16)     &   0.94          \\
0.4   & 0.31671(42)   &	0.1208(97)    &   0.67          \\
0.6   & 0.28996(26)   &	0.0706(57)    &   0.36          \\
0.8   & 0.26838(16)   &	0.0415(36)    &   0.19          \\
1.0   & 0.24988(10)   &	0.0236(23)    &   0.10          \\
2.0   & 0.182623(10)  &	-0.00389(11)  &   $ < 10^{-2}$  \\
4.0   & 0.112434(18)  &	-0.00544(43)  &	  0.02          \\
6.0   & 0.079244(17)  &	-0.00399(39)  &	  0.03          \\
8.0   & 0.060735(14)  &	-0.00307(33)  &	  0.03          \\
\hline
\end{tabular}
\caption{Results for the fit of the function $g$ for $N \ge 4$. $a$ is the leading term in the large-$N$ limit, while $b$ is the coefficient of the ${\cal O}(1/N^2)$ correction. The reduced $\chi^2$ is also listed.}
\label{tab:condensate:g}
\end{center}
\end{table}

\end{document}